\documentclass[prb,twocolumn,showpacs,aps,floats,floatfix,superscriptaddress]{revtex4}

\usepackage{graphicx}

\usepackage[dvips]{xcolor}

\begin{document}
\title{Graphene on Ir(111) surface: From van der Waals to strong bonding}
\author{R. Brako}
\author{D. \v Sok\v cevi\'c}
\affiliation{Rudjer Bo\v{s}kovi\'{c} Institute, 10000 Zagreb, Croatia}
\author{P. Lazi\'c}
\author{N. Atodiresei}
\affiliation{Institut f\"ur Festk\"orperforschung (IFF), Forschungszentrum J\"ulich, 
52425 J\"ulich, Germany}
\date{\today}

\begin{abstract}
We calculate the properties of a graphene monolayer on the Ir(111) surface, using 
the model in which the periodicities of the two structures are assumed equal, instead of 
the observed slight mismatch which leads to a large superperiodic unit cell. 
We use the Density Functional Theory approach supplemented by the recently 
developed vdW-DF nonlocal correlation functional. The latter is
essential for treating the van der Waals interaction, which is crucial for the
adsorption distances and energies of the rather weakly bound graphene. When 
additional iridium atoms are put on top of graphene, the electronic 
structure of C atoms acquires the $sp^3$ character and strong bonds with 
the iridium atoms are formed.
We discuss the validity of the approximations used, and the relevance for
other graphene-metal systems.
\end{abstract}

\maketitle

\section{Introduction}
\label{sec:introduction}

Graphene is a one-atom thick two-dimensional structure of carbon atoms arranged 
in a honeycomb lattice. It is (conceptually at least)
at the origin of all other graphitic forms,~\cite{Geim2007} 
including the three-dimensional
graphite, one-dimensional carbon nanotubes and zero-dimensional fullerenes. 
The planar geometry and the exceptional strength of graphene~\cite{Lee2008} are due to 
the $sp^2$ bonds between atoms. Single-layer graphene has been obtained 
by micromechanical cleavage of graphite and by growth on SiC and metal surfaces. The recent
large increase of interest in graphene is due both to the theoretical implications of 
its unique electronic properties and to its potential applicability, in particular as a novel 
material for electronics.

As the building block of graphite and as the adsorbate on many surfaces, 
graphene bonds to its surroundings 
only weakly, and the character of the bonding is largely van der Waals (vdW). 
Occasionally, stronger bonding occurs 
without destroying the geometry of the graphene lattice,~\cite{Wintterlin2009}
for example on Ni(111)~\cite{Ni111} and 
Ru(0001)~\cite{Ru0001Marchini,Ru0001Martoccia,Ru0001Moritz} surfaces.
Graphene on Ir(111) is an example where, 
depending on conditions,
both kinds of bonding can occur. Monolayer 
graphene is vdW physisorbed, and the characteristic graphene lattice can be 
clearly seen in STM images,~\cite{NDiaye2006}
while with additional Ir clusters on top the carbon-metal bonds become stronger.

Large-cell Density Functional Theory (DFT) calculations of graphene on Ir(111) using PBE GGA 
(Ref.~\onlinecite{NDiaye2006}) and LDA
(Ref.~\onlinecite{Feibelman2008}) have been performed. 
However, experiments and calculations reveal a subtle interplay between
van der Waals bonding and stronger electronic interaction with the substrate. The
vdW interaction is not properly described in the standard local (LDA) and
semilocal (GGA) DFT functionals, which provides a serious obstacle
to the complete understanding of the nature of the bonding.

In this paper we apply the recently developed extension to the DFT, which 
replaces the semi-local (i.e. depending upon the gradient of the electronic density)
correlation term with a fully non-local one (depending upon the electronic densities
at different points in space), which can describe the van der Waals forces even between
two fragments of matter with non-overlapping electronic densities.
Due to computational complexity we had to 
abandon the large supercell which aims to describe more realistically
the graphene and Ir(111) surface with their slightly different atomic periodicities,
and opt for an approximate description by a smaller commensurate unit cell.
While the quantitative accuracy of the results suffers (but we argue that it is 
a quite limited and controlled problem), the approximation makes the transition from 
weak to strong bonding easier to analyze and understand.

\section{Graphene Binding in Graphite}
\label{sec:graphite}

\begin{figure}[ht!]
\begin{center}
\resizebox{0.85\columnwidth}{!}{
\includegraphics[width=0.4\textwidth,clip=true]{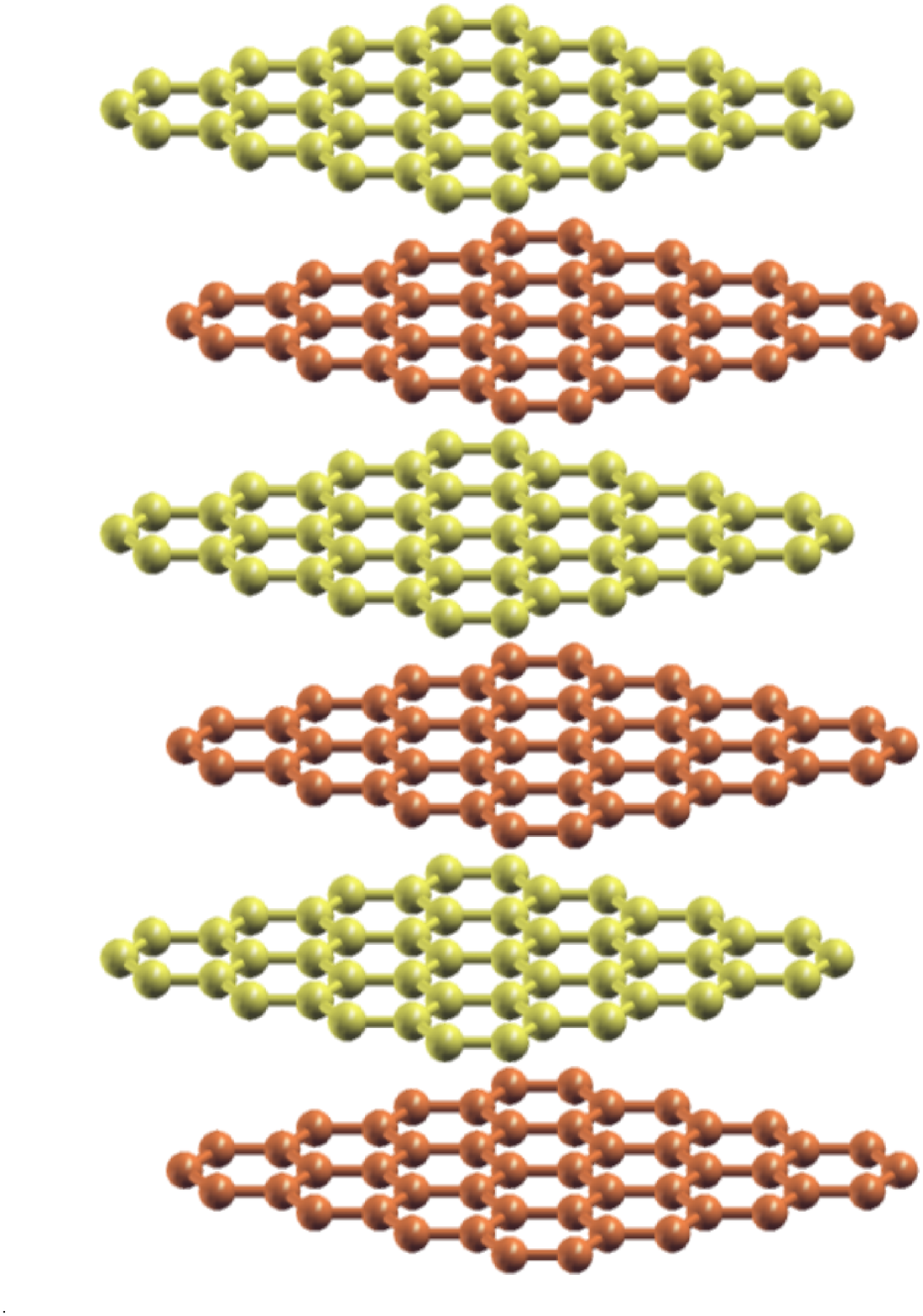}
\hspace{0.08\textwidth}
\raisebox{0.2\textwidth}{\includegraphics[width=0.35\textwidth,clip=true]{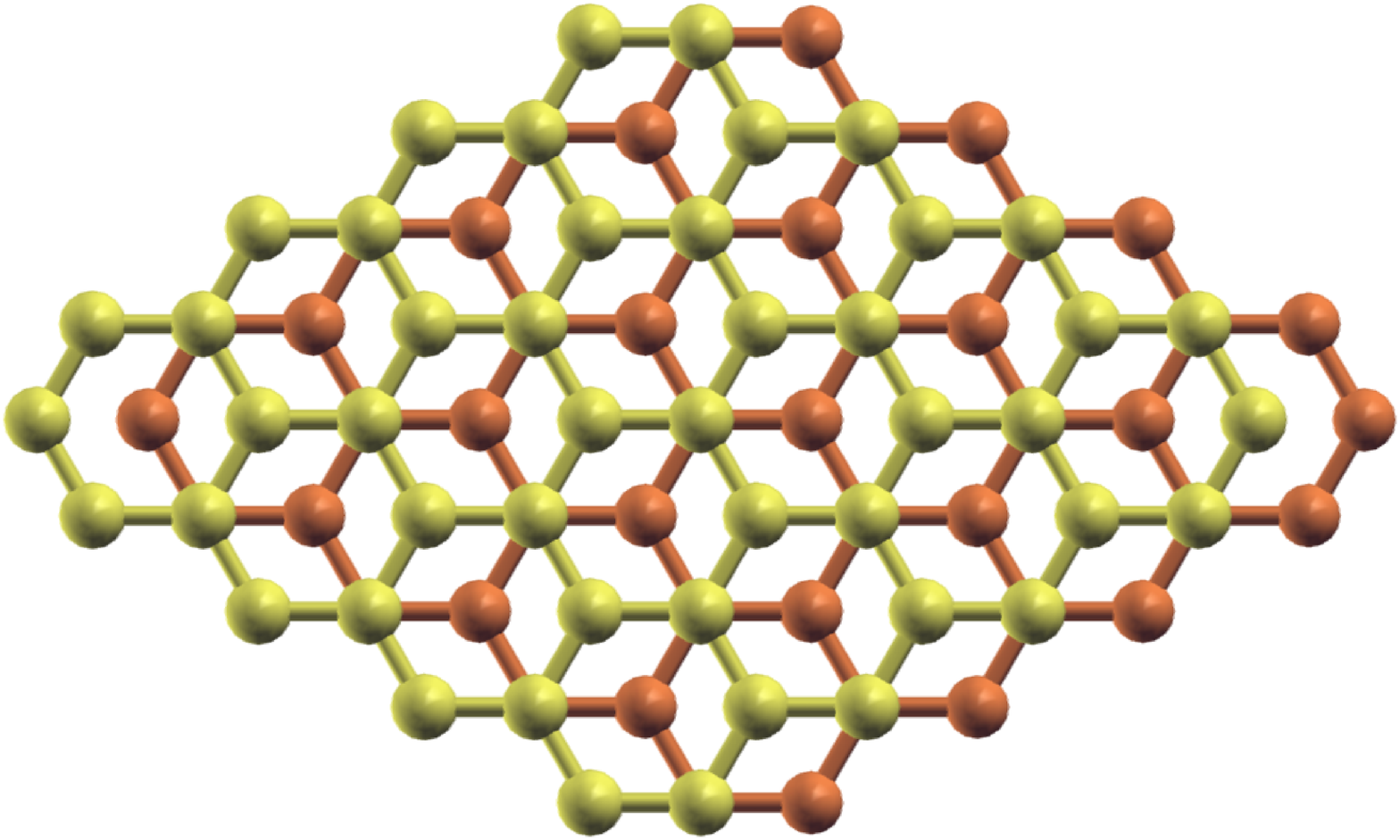}}
}
\caption{\label{fig:graphene6}
The structure of graphite, side and top view. The layers are stacked in AB 
order, so that half of the C atoms lie 
in chains along
the direction perpendicular to the graphene planes, while the
other half alternates in the other two high-symmetry positions.
The periodicity in the
perpendicular direction is $c$, i.e. the interlayer distance is $c/2$.
For clarity, $c$ has been exaggerated by about a factor of 3.
}
\end{center}
\end{figure}
\begin{figure}[ht!]
\rotatebox{0}{
\resizebox{0.90\columnwidth}{!}{
\includegraphics[clip=true]{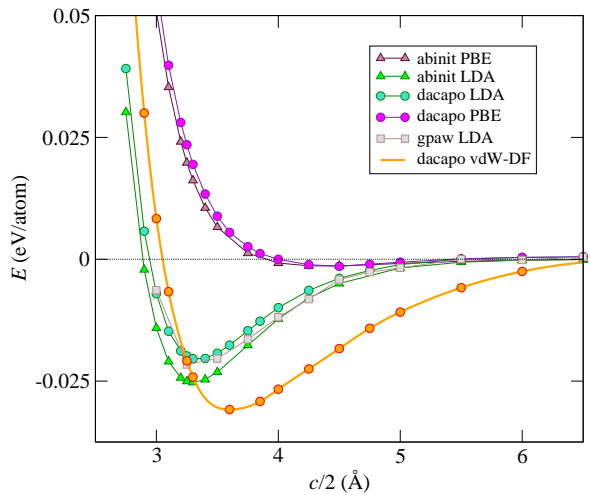}
}}
\caption{\label{fig:graphite}
Binding energies of graphite AB structure as a function of interlayer distance: 
Comparison of various DFT codes, plane-wave based dacapo~\cite{dacapo} and 
abinit,~\cite{abinit} and real-space gpaw.~\cite{gpaw} Local LDA and
semilocal PBE GGA results are shown. The curve labeled vdW-DF is the energy when the PBE GGA
correlation has been replaced by a fully non-local correlation~\cite{Dion2005} 
using the JuNoLo numerical code~\cite{JuNoLo}.
}
\end{figure}

We have first applied our methods to 
graphite, i.e. graphene sheets arranged in the most stable AB stacking,
shown in Fig.~\ref{fig:graphene6}.
This is a much studied system with good experimental data and calculated values
of structural and energetic parameters. It has the same complexities of having both 
the weak vdW and strong chemical bonds as our principal subject of interest,
graphene on Ir(111).
A single sheet of graphene presents no difficulties for the standard 
DFT GGA approach, giving the C-C distance of 1.42~{\AA}, corresponding
to a lattice constant of 2.46~{\AA}, in agreement with the experiment.
Next, we performed the Density Functional calculations of stacks of 
graphene sheets using
several flavors of LDA and GGA, implemented in various numerical codes. 
The calculated cohesive energies are shown in Fig.~\ref{fig:graphite}, as a
function of the interlayer separation.
After that we used the nonlocal vdW-DF functional for the 
correlation.~\cite{Dion2005,Langreth2009}

The pure DFT results agree well for all programs used, and are even quite
insensitive upon the (lack of) full self-consistency. 
For example, dacapo calculations use the PW91 GGA functional, but the dacapo LDA curve 
shown in the figure, obtained 
evaluating the LDA functional on the electron density calculated with PW91,
agrees quite well with the fully
selfconsistent abinit LDA results. 
GGA calculations give little or no bonding, and 
LDA gives (apparently) reasonable bonding energies and distances, comparable
to experimental values. The reason for
the failure of GGA is intuitively clear: The semilocal gradient approximation 
cannot describe well the inherently non-local van der Waals interaction, which
exists even between subsystems with completely non-overlapping electronic densities.
The apparent success of LDA is somewhat perplexing, since it is even more 
local than the more advanced GGA, the latter indeed being more successful when it 
comes to chemically bound systems. There are strong indications that the 
agreement of the LDA results is largely fortuitous, as discussed further on.

We have further investigated the problem by applying the 
vdW-DF theory,~\cite{Dion2005} which is at present the most promising 
approach for treating the non-local correlation.
It consists in replacing the semilocal (gradient) part of the 
GGA correlation functional by a fully non-local term, which still depends only
upon the electronic density, in the true spirit of the Density Functional theory. 
We have applied vdW-DF 
as implemented in the JuNoLo code~\cite{JuNoLo}
in a post-processing approach, i.e. we used the electron 
densities obtained in the standard GGA calculation to evaluate the non-local 
vdW-DF correlation, and inserted it into the total energy instead of
the semilocal correlation. This
approach is, of course, not fully selfconsistent, since the DFT potential and
the Kohn-Sham wavefunctions, and
hence the electron density can depend upon the details of the correlation functional used.
However, a recent selfconsistent implementation of the vdW-DF correlation 
functional shows that the differences are negligible.~\cite{Thonhauser2007} We 
have therefore relied on the post-processing approach which is less time consuming and
avoids any intervention in the code of the DFT programs. 
Changing the correlation contribution changes the total energy, and
therefore the forces acting on the atoms as well. However, all atomic configurations
which we consider here have high symmetry, where we can sweep the interesting range of
interlayer separation ``by hand'' in order to find the optimum configuration. 
The lack of fully selfconsistent 
atomic relaxations inherent to such an approach is not a major problem, as discussed 
later on. We furthermore note that we have not followed the suggestion 
put forward by the authors of the vdW-DF theory
to use the revPBE exchange functional,\cite{Dion2005} and have instead continued 
using the PBE exchange. 
Although revPBE exchange seems to compensate for too large binding energies
for several van der Waals bound systems, it gives worse equilibrium distances,
and the same improvement does not seem to occur in cases of strong bonding.

The vdW-DF results shown by a thick line in Fig.~\ref{fig:graphite} 
are qualitatively similar 
to the LDA results, but with some important differences. The equilibrium 
distance at around 3.6~{\AA} is 
larger than the LDA result, as is the binding energy of 30.8~meV per carbon atom.
The vdW-DF  attractive potential has clearly a longer range than LDA,
which reflects the long-range nature of the van der Waals attraction
and reveals the fortuitous character of the agreement with LDA around the minimum.
The vdW-DF values for the interlayer separation and interaction energy
agree well with recent experimental results
and theoretical calculations (See Ref.~\onlinecite{Spanu2009} and references therein).
Recently, it has been found that the behavior of the nonretarded van 
der Waals interaction between nonoverlapping anisotropic nanostructures that have 
a zero electronic energy gap should be different than predicted from the 
the usual sum of $R^{-6}$ contributions,~\cite{Dobson} but this is probably 
relevant only in the extreme asymptotic regime.

We conclude
that the inclusion of the van der Waals interaction 
is essential to reproduce physical properties of graphite, and that the vdW-DF
approach successfully treats all aspects of the graphene binding in graphite.

\section{Structure of graphene on I\lowercase{r}(111) surface}
\label{sec:graphene-ir}
\begin{figure}[ht!]
\rotatebox{0}{
\resizebox{0.80\columnwidth}{!}{
\includegraphics[clip=true]{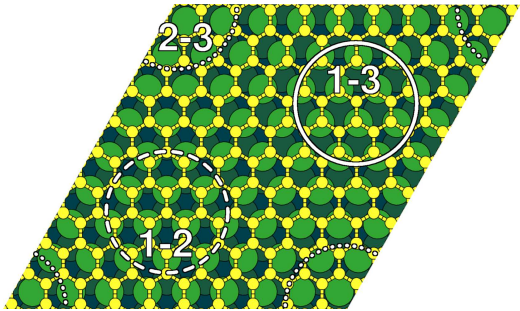}
}}
\caption{\label{fig:moire}
Moir\'e superstructure of $10 \times 10$ graphene on $9 \times 9$ Ir(111)
unit cell. 
The C atoms 
are (approximately) above the first and third layer Ir atoms 
within the circle labeled 
1-3,
above the first and second layer in 
1-2,
and above the second and third layer
in 
2-3.~\cite{123}
}
\end{figure}
Graphene monolayers of high structural quality, extending over tens of nanometers 
and even up to micrometer size, orientationally well aligned with the substrate,
have recently been obtained by hydrocarbon decomposition on Ir(111).~\cite{NDiaye2006}
The lattice constants of graphene and the Ir(111) surface differ at room 
temperature by around 10\%, and the STM micrographs clearly show the moir\'e pattern
due to the lattice mismatch. In Fig.~\ref{fig:moire} the supercell with a 
$10 \times 10$ graphene lattice on top of $9 \times 9$ structure of iridium atoms is 
shown. A further intriguing feature is observed when additional iridium atoms are 
adsorbed on top of the graphene. STM images show that the adatoms form regular
arrays on clusters, selectively bound to certain
regions of the moir\'e pattern.~\cite{NDiaye2006,Feibelman2008}

Density functional calculations 
employing the PBE GGA functional~\cite{NDiaye2006} and LDA functional~\cite{Feibelman2008} 
on a supercell similar to the one in Fig.~\ref{fig:moire}
have been performed. It has been found thet the PBE GGA functional gives 
almost no bonding of graphene monolayer
on Ir(111) (Refs.~\onlinecite{NDiaye2006} Erratum, \onlinecite{Lacovig2009}), while the LDA functional
gives reasonable results for bonding energies and interatomic distances, both without
and with additional clusters on top.~\cite{Feibelman2008} These results are
qualitatively reminiscent of our results for graphite, i.e.
we again see an apparent success of the quite basic LDA, which signals that a 
more detailed investigation is necessary.

Our strategy is similar to the approach used for graphite in Section~\ref{sec:graphite}.
We start with the standard DFT and later investigate the effects of the non-local 
correlation. We do not use the realistic large unit cell shown in Fig.~\ref{fig:moire},
but instead we compress the Ir(111) substrate in the surface plane (coordinates $x, y$)  
so that it matches the lattice constant of graphene. By changing the 
phase of the carbon atoms with respect to the underlying lattice of Ir atoms, we are able to
simulate
(approximately) any point in the supercell in Fig.~\ref{fig:moire}.
We have done most calculations for the region labeled 
1-3,
which both
experiment and calculations show to be the most strongly bonding, with only a few
checks of the other regions.

Calculations of commensurate graphene-metal surface systems has been
done for several metal surfaces, either by adjusting the substrate
lattice constant~\cite{Giovannetti2008} or the graphene lattice 
constant~\cite{Vanin}. We shall discuss these calculations into more 
detail later on.

The mismatch of the lattice constant of graphene (2.46~{\AA}) and that of the
Ir(111) surface (2.73~{\AA}, corresponding to the conventional fcc lattice
constant $a_0=3.86$~{\AA}) is around 10\%, clearly 
larger than those in Ref.~\onlinecite{Giovannetti2008}, which are 
in the range of 0.8-3.8\%.
In order to minimize possible artefacts due to the squeezing of the iridium 
substrate to fit the graphene lattice we optimized the lattice constant 
of the iridium substrate in the $z$ direction. To that end, we performed 
calculations of iridium bulk with compressed (111) planes and
allowed it to relax in the perpendicular direction. The lattice constant 
in the $z$ direction increased by about 10\% to 4.24~{\AA}, and we used this 
lower-symmetry iridium lattice, compressed in $x-y$ directions and expanded 
in $z$, as the substrate in our calculations. 
From the point of view of quantitative accuracy, a  calculation using a
large supercell and ``natural'' iridium substrate would be preferred,
but our approach
enables a clear insight into the bonding properties,
which would be at risk to remain buried and hard to see in a more realistic
large calculation. 

Another important aspect of graphene interaction with the Ir(111) surface
can be inferred from the band structure of Ir(111) along high-symmetry
directions of the surface Brillouin zone. Our calculations based on the
DFT Kohn-Sham eigenstates~\cite{Pletikosic10}
as well as ARPES experiments~\cite{Pletikosic2009,Pletikosic10} 
show that there is an energy gap around the $K$ point of the surface Brillouin
zone, extending 
from just below the Fermi level down to almost 1.5~eV binding energy. 
The vertex of the ``Dirac cone'' of the $\pi$ bands of graphene adsorbed on 
Ir(111) lies entirely  within this gap.~\cite{Pletikosic2009} 
The weak interaction of graphene with 
the iridium substrate can be attributed to this mismatch of the electronic 
states, since the unsaturated $\pi$ bands of the Dirac cones don't
have any substrate states with the same 
momentum $k$ and energy $E$
to hybridize with.
We have also checked the band structure of our compressed iridium surface and
found that the band gap around the $K$ point is still present and has a similar 
shape, which implies that the weak character of the graphene interaction with
Ir(111) will not be much affected by the use of the compressed substrate.

In our DFT calculations we use a three-layer Ir(111) slab with the adjusted 
lattice constants in the $x-y$ and the $z$ directions as explained above. 
It would have been easy to use  a thicker substrate,
but this would add no further quantitative accuracy, considering 
other
simplifications and approximations used.

\section{
DFT calculations of graphene on I\lowercase{r}(111) and 
I\lowercase{r}-graphene-I\lowercase{r} sandwiches }
\label{sec:sandwiches}

\begin{figure}[ht!]
\resizebox{0.98\columnwidth}{!}{
\includegraphics[width=0.170\textwidth,clip=true]{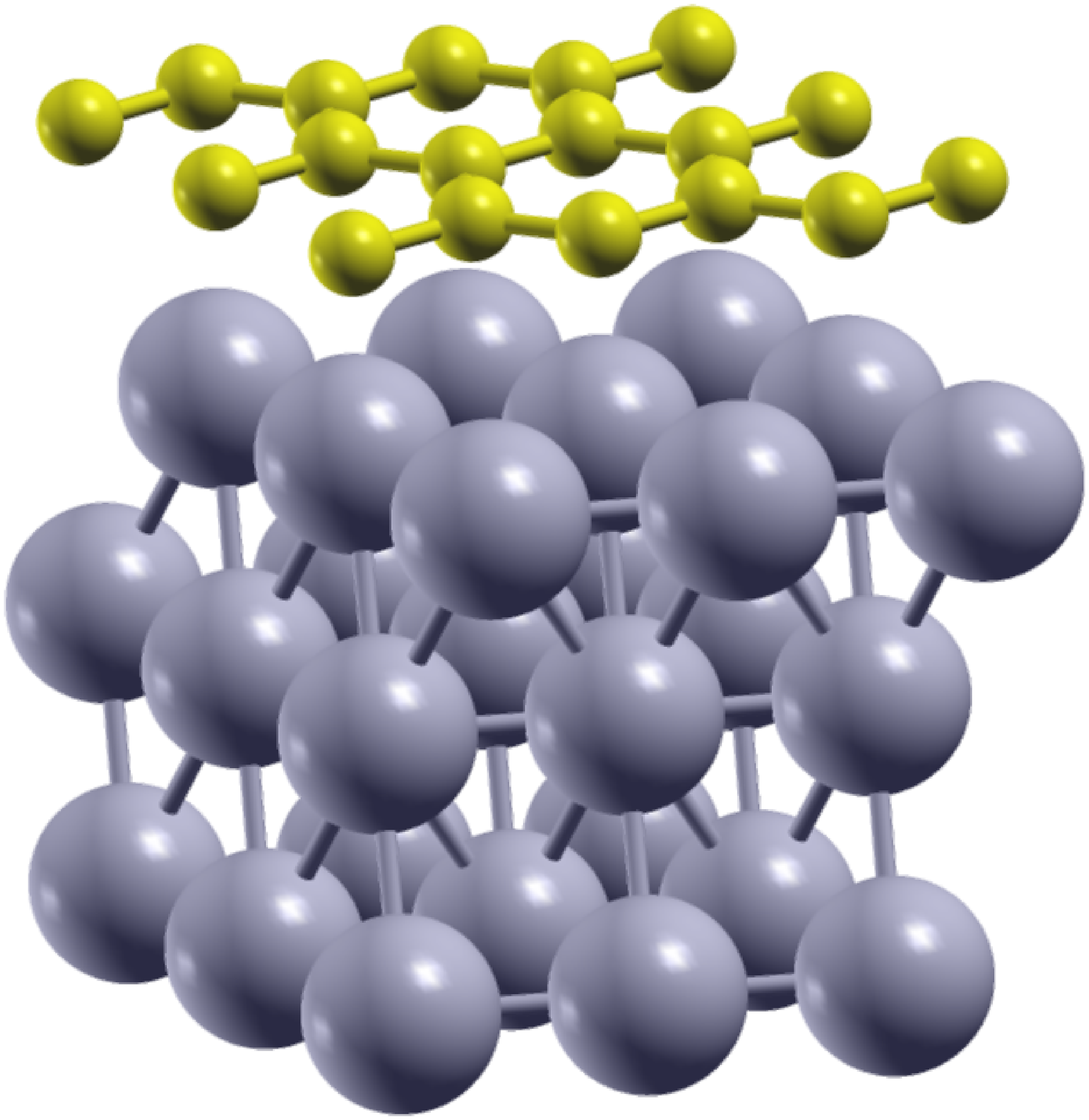}
(a)
\hspace{0.05\linewidth}
\includegraphics[width=0.08\textwidth,clip=true]{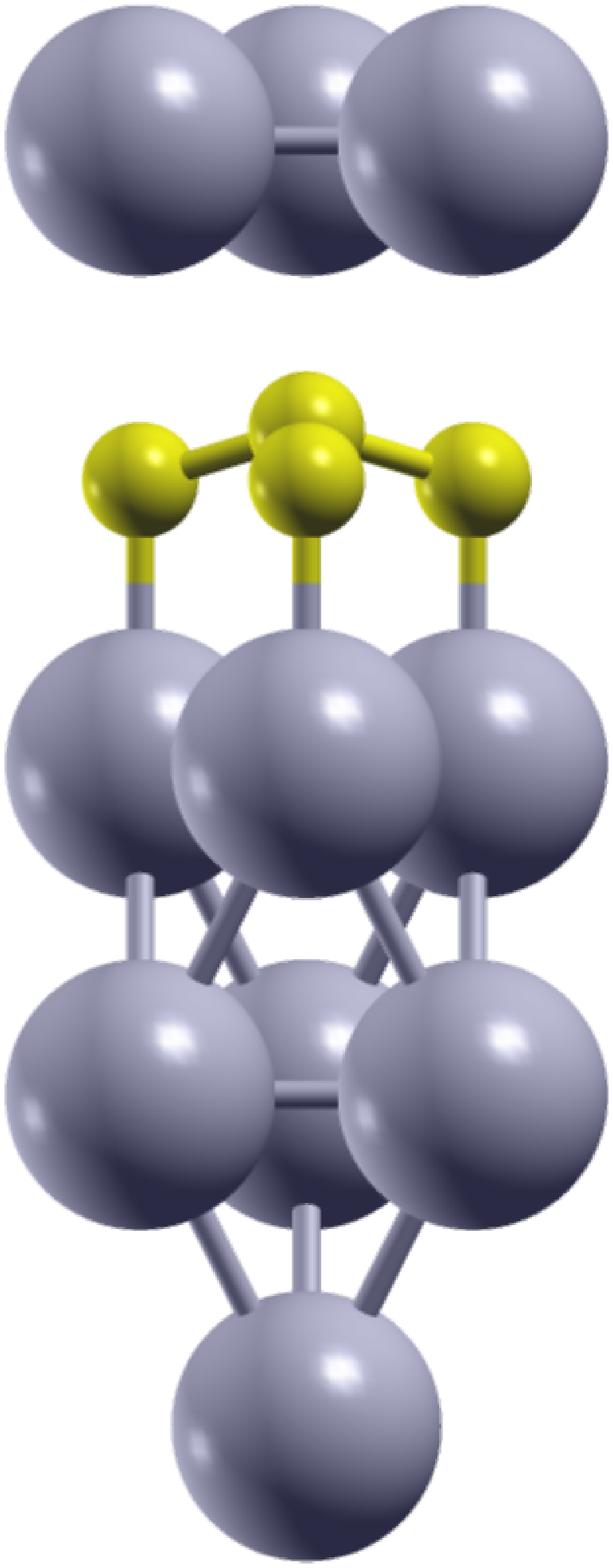}
(b) 
\hspace{0.05\linewidth}
\includegraphics[width=0.08\textwidth,clip=true]{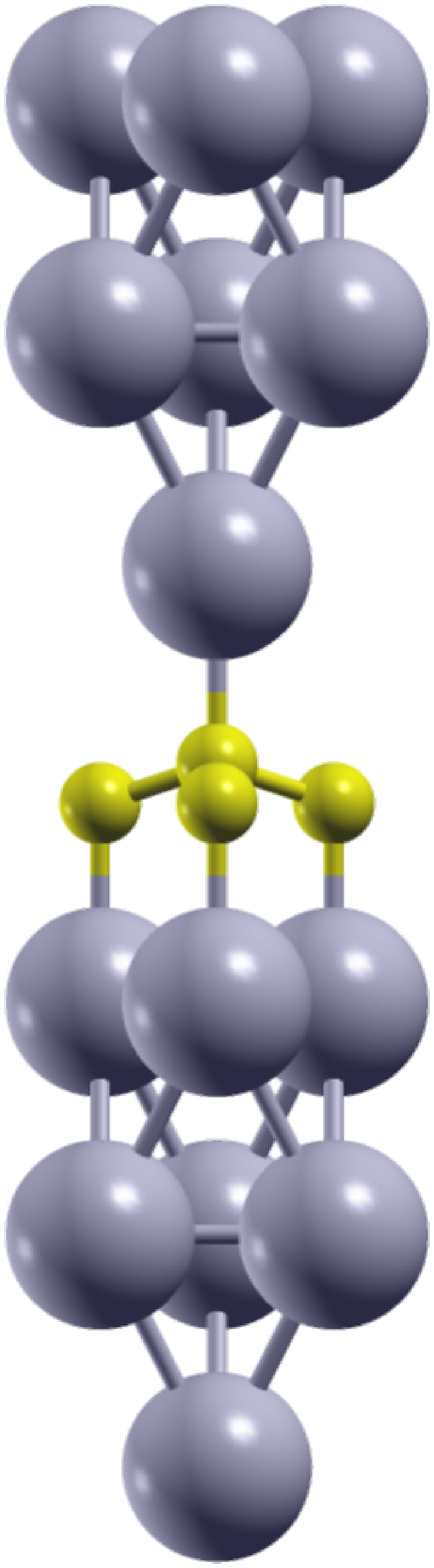}
(c)
\hspace{0.05\linewidth}
\includegraphics[width=0.12\textwidth,clip=true]{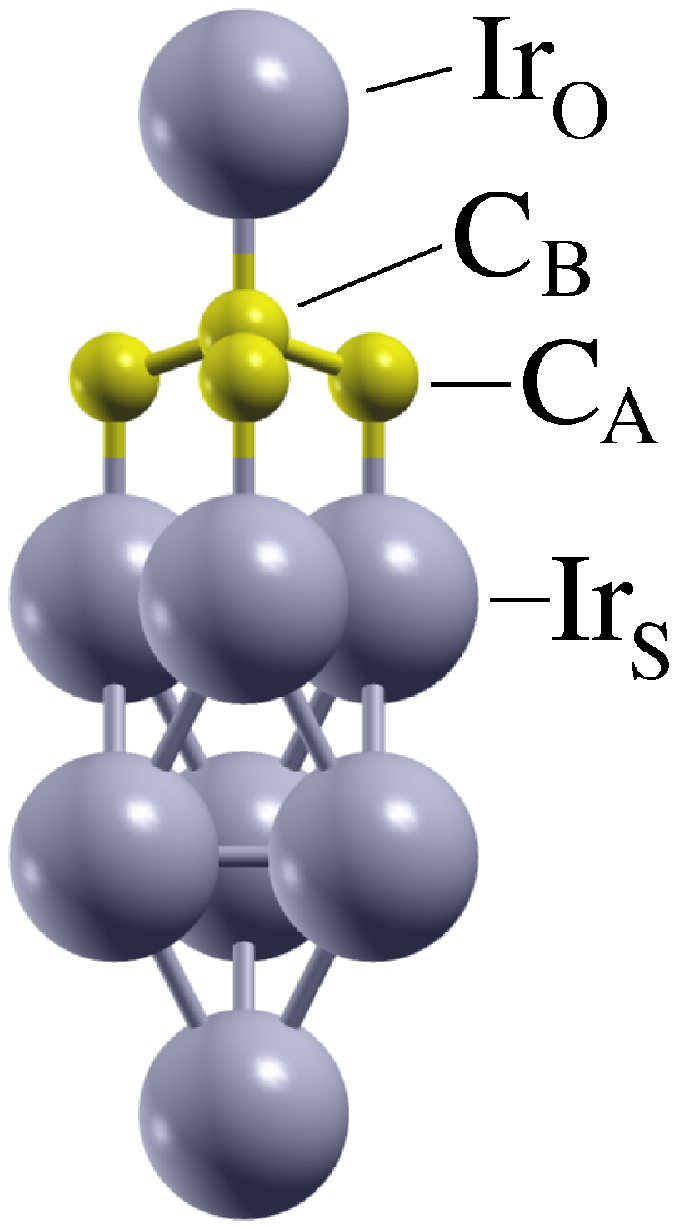}
(d)
}
\caption{\label{fig:sandwiches}
The four structures considered in the paper. For clarity, only a few atoms from each
atomic layer are shown in structures (b)-(d).
}
\end{figure}

For our calculations
we have chosen four characteristic structures of graphene on iridium, 
shown in Fig.~\ref{fig:sandwiches}.
The structures are periodic in the $x-y$ plane and extend to infinity, 
but for clarity we show only a small symmetric cluster of atoms for each one.
There are 
two atoms in the unit cell of graphene,
which in the following we denote 
by C$_\mathrm{A}$ and C$_\mathrm{B}$, as illustrated in Fig.~\ref{fig:sandwiches}~(d).
We denote the iridium atoms in the first substrate 
layer by Ir$_\mathrm{S}$, and the atoms in the first overlayer as 
Ir$_\mathrm{O}$.~\cite{ABnotation}

The structures were chosen so that they illustrate Ir-C bonds of various 
character, with overall bonding strength increasing from structure (a) 
to structure (d). The iridium substrate is modeled by three atomic layers
in all cases.
The structures are:
(a) Monolayer graphene on Ir(111) with C$_\mathrm{A}$ above Ir$_\mathrm{S}$
and C$_\mathrm{B}$ above third layer Ir, which illustrates the most stable
regions of the moir\'e pattern of a monolayer graphene on Ir(111); 
(b) Graphene monolayer as in (a), but with a single overlayer of iridium 
atoms, with Ir$_\mathrm{O}$
located above the centers of the hexagonal rings of the graphene;
(c) Three additional layers of iridium, with Ir$_\mathrm{O}$ above 
C$_\mathrm{B}$; 
(d) A single Ir overlayer, with atoms in the same positions as in the first 
layer in (c). In (c) and (d) there is one iridium atom below C$_\mathrm{A}$
and another one above C$_\mathrm{B}$, which models the geometries of the 
stable iridium clusters on top of graphene observed in the experiment. 

\subsection{Standard DFT only}
\label{sub:DFT}

\begin{figure}[ht!]
\rotatebox{0}{
\resizebox{0.90\columnwidth}{!}{
\includegraphics[clip=true]{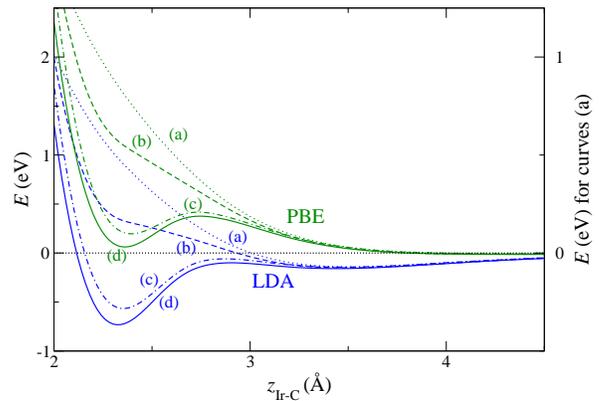}
}}
\caption{\label{fig:LDA_PBE}
DFT (LDA and PBE GGA) energies for graphene on Ir, structure (a) in 
Fig.~\ref{fig:sandwiches}, and Ir-graphene-Ir sandwiches, structures
(b)--(d) in Fig.~\ref{fig:sandwiches}.
In this and the following graphs the energies are given per unit cell, i.e.
two C atoms and one Ir atom in each iridium layer.
}
\end{figure}

We first calculated the dependence of the interaction energy upon Ir-graphene 
separation for structures (a)-(d), using the standard LDA and GGA PBE 
functionals. 
The results are shown in Fig.~\ref{fig:LDA_PBE}. Two comments are in order. For 
structure (a), the energy scale in Fig.~\ref{fig:LDA_PBE} is smaller by 
a factor of two, since the graphene has 
iridium atoms only on one side, and the interaction (in particularly the 
repulsion at small distances) is expected to scale with the number of
neighboring atomic planes. Secondly, for the sandwich structures (b)-(d) the graphene
layer was at the beginning of the calculations
placed symmetrically between the nearest Ir layers, each of 
them at a distance of $z_\mathrm {Ir-g}$, but was allowed to
relax in the course of the calculation. At small separations $z_\mathrm {Ir-g}$
the graphene buckles, with C$_\mathrm{A}$ atom moving towards Ir$_\mathrm{S}$
and C$_\mathrm{B}$ towards the overlayer.
This is not a major problem, since the two atoms move by almost the same
amount in opposite direction,
even in the case of the less symmetrical structure (b), so that $z_\mathrm {Ir-g}$
still measures the $z$-averaged position of the graphene plane. 
There is no buckling at $z_\mathrm {Ir-g}$ separations larger than say 3~{\AA}. Since the
relaxation was done according to the forces calculated in GGA functional, 
which does not develop any appreciable attractive potential well at these
distances, there was no
danger that the graphene layer would be attracted to the iridium atoms on
one side. Iridium atoms were not relaxed.

The results for all structures almost coincide for $z_\mathrm {Ir-g}$ larger
than around 3.3~{\AA}, for both GGA and LDA, i.e. the interaction at
physisorption distances does not show a large corrugation along the $x-y$
coordinates. The sandwich 
structures (b)-(d) show the tendency to form a strong bond at small graphene-Ir
distances. The minimum is around 2.3~{\AA}, and is quite sensitive on the
relative position of the atoms in the $x-y$ plane.
Thus in the unfavorable structure (b), where the Ir atoms in the additional layer 
do not lie directly above the C atoms, there is only a kink in the 
interaction energy at about 2.3~{\AA}, hinting that there is a tendency towards
strong chemical bonding. The structures (c) and (d) develop a distinct potential well
around that distance, which is not deep enough in the PBE GGA functional
calculation, but with the LDA functional it becomes the stable configuration
with more than 0.5~eV binding energy. Note that the quantity $z_\mathrm {Ir-g}$
measures the distance to the average $z$ coordinate of the graphene layer, and 
since there is a buckling of about 0.2~{\AA} of C atoms towards the nearest
Ir atom, the Ir-C distance is actually around 2.1~{\AA}, as discussed more into
details later on.

These results 
prompt us to reevaluate even the standard DFT calculations
in the region of strong bonding at small Ir-C distances, where
the graphene layer significantly changes its electronic character.
Up to now
we have consistently used the lattice constant of free graphene, assuming
that it is optimal for the bound system too (and we went to the trouble of 
compressing the Ir layers accordingly). This may not be true in the region of 
strong bonding, and we first check this.

To that end, we performed standard DFT calculations of structures (c) and (d) with the 
lattice constant in the $x-y$ plane slightly expanded  from the value of free 
graphene (3.478~{\AA}, C--C distance 1.42~{\AA}) and accordingly reduced in the $z$
direction, and checked whether there was any improvement of
the total energy. In order to evaluate the bonding energy we also had to calculate
the energy of separated Ir and graphene slabs (corresponding to 
$z_\mathrm {Ir-g} \rightarrow \infty$) for each value of the expanded lattice
constant, and subtract it from the energy of the interacting system.
In Fig.~\ref{fig:vdW-DF} we show the results for structures (c) and (d). The
unconnected circles are the non-optimized results for the two structures taken from 
Fig.~\ref{fig:LDA_PBE}, while the connected circles are the best results
obtained by expanding the lattice. We see that the energy improves 
significantly in the region of strong bonding, where the optimum lattice constant
at the position of the minimum, $z_\mathrm {Ir-g} \sim 2.3$~{\AA},
increases from the free graphene value of  3.478~{\AA} to 3.65~{\AA} for 
structure (c) and to 3.72~{\AA} for structure (d). The fact that 
in the region of weak binding, for $z_\mathrm {Ir-g} > 3$~{\AA},
there is no improvement of energy and the optimal lattice constant remains 
at the value of free graphene (i.e. the large graphene stiffness dominates
the energy balance) indicates that the procedure of optimizing
the lattice constant is consistent with other 
approximations used in the calculations.

\subsection{DFT with vdW-DF}
\label{sub:vdW-DF}

\begin{figure}[ht!]
\resizebox{0.80\columnwidth}{!}{
\includegraphics[clip=true]{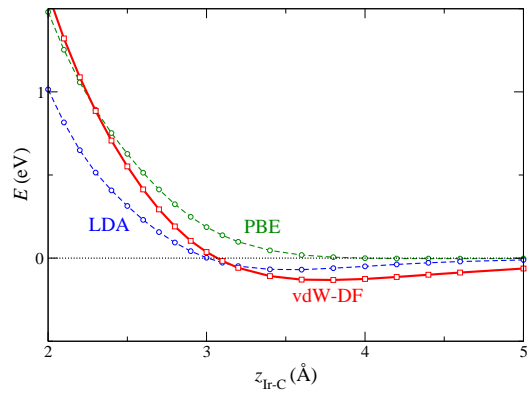}
}
\caption{\label{fig:vdW-DF-1}
Standard DFT (LDA and PBE GGA) and vdW-DF energies for graphene monolayer
on Ir(111), structure (a) in Fig.~\ref{fig:sandwiches}.
}
\end{figure}

\begin{figure}[ht!]
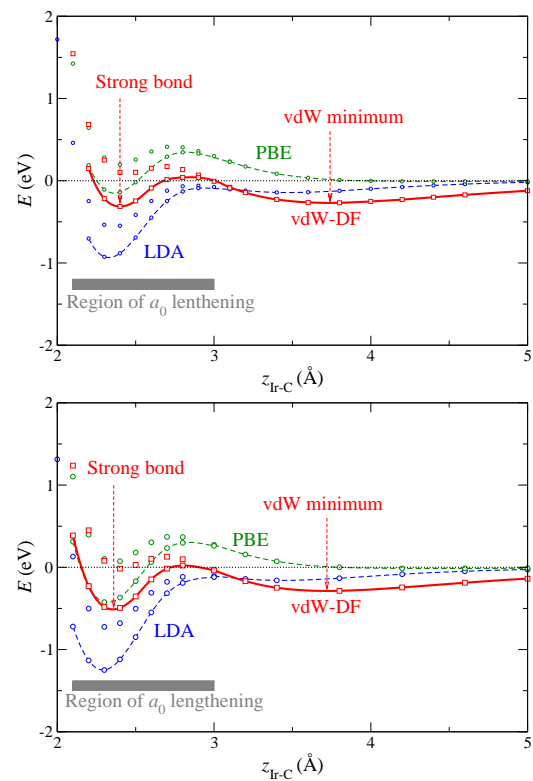

\rotatebox{0}{
\resizebox{0.80\columnwidth}{!}{
\includegraphics[clip=true]{Etot-NL-c.eps} 
}}
\rotatebox{0}{
\resizebox{0.80\columnwidth}{!}{
\includegraphics[clip=true]{Etot-NL-d.eps}
}}
\caption{\label{fig:vdW-DF}
Standard DFT (LDA and PBE GGA) and vdW-DF energies for Ir-graphene-Ir sandwiches,
structures (c) and (d) in Fig.~\ref{fig:sandwiches}. The lines connect points
for which the energy has minimum when allowing the structures to slightly
expand laterally, as explained in the text.
}
\end{figure}

In order to account for the van der Waals interaction and the effects of the
long-range correlation in general, we 
applied the vdW-DF approach in a post-GGA procedure to DFT results for the 
structures (a), (c) and (d). Here the full power of the vdW-DF 
correlation approach becomes obvious, because due to its ``seamless'' character we
can apply it at all graphene-Ir distances, i.e. at all coupling strengths,
without worrying that it may spoil the GGA results which are good for strong 
bonding.~\cite{puzzle} 

The results for structure (a) are shown by squares in Fig.~\ref{fig:vdW-DF-1},
where the PBE and the LDA results are the same as in Fig.~\ref{fig:LDA_PBE}, and
the energy calculated using vdW-DF is shown by a thick line and squares.
A clear van der Waals attractive well develops, deeper than the shallow
well in the LDA calculation, and with the minimum at a larger graphene-substrate
distance, around 3.7~{\AA}. 
Fig.~\ref{fig:vdW-DF} shows 
similar results for structures (c) and (d).
The unconnected points are for the lattice constant of 
free graphene, as in Fig.~\ref{fig:LDA_PBE}, 
and the points connected by lines for the optimized
expanded lattice constant. 
The van der Waals potential well is similar to Fig.~\ref{fig:vdW-DF-1} but 
approximately two times deeper (note the different scale on the energy axis), 
since the graphene interacts both with the 
iridium substrate and overlayer.
The depth and shape of the chemisorption minimum at around 2.3~{\AA}
is less affected, but the barrier between the two minima is much decreased
compared with the DFT results.

\subsection{Discussion of the results}
\label{sub:discussion}

Detailed information about the nature of C--C and C--Ir bonds can be inferred
by examining the geometry of the graphene lattice around the minima of the 
interaction energy in Fig.~\ref{fig:vdW-DF}.
At the physisorption minimum, $z_\mathrm {Ir-g} \sim 3.7$~{\AA}, the graphene 
lattice is perfectly planar, and the graphene stays at midpoint between two
neighboring iridium layers. 
The same is true
for all structures in Fig.~\ref{fig:sandwiches}, 
and in fact for other geometries 
such as monolayer graphene over the 
1-2 and 2-3 regions in 
Fig.~\ref{fig:moire}. Due to the smoothness of the potential with respect to
the translation of graphene along the surface, graphene flakes 
physisorbed on Ir(111) are quite mobile, both translationally and rotationally,
which is an important mechanism in aggregation and growth of large graphene 
islands.~\cite{Coraux2009}

\begin{table}[ht]
\begin{centering}
\begin{tabular}{c c c c c c c}
\hline
\makebox[4em]{Structure} &
\makebox[3em]{$a_0$} &
\makebox[3em]{$z_{\mathrm{S-A}}$} &
\makebox[3em]{$z_{\mathrm{buck}}$} &
\makebox[3em]{$z_{\mathrm{B-O}}$} &
\makebox[3em]{$d_{\mathrm{A-B}}$} &
\makebox[3em]{$\alpha$} \\
\hline
(c) & 3.65 & 2.17 & 0.40 & 2.17 & 1.54 & 105$^\circ$ \\
(d) & 3.72 & 2.22 & 0.41 & 2.17 & 1.57 & 105.3$^\circ$ \\
\hline 
\end{tabular}
\caption{
Bond length and angles at the strong bonding energy minima of structures (c) and
(d). All lengths are in {\AA}. Here $a_0$ is the optimal lattice constant of
the structure in the $x-y$
plane, slightly larger than the free graphene value of 3.478~{\AA}, $z_{\mathrm{S-A}}$
is the distance between the substrate atom Ir$_\mathrm{S}$ and C$_\mathrm{A}$,
$z_{\mathrm{buck}}$ is the buckling of graphene, i.e. the difference of $z$ 
coordinates of atoms C$_\mathrm{A}$ and C$_\mathrm{B}$, $z_{\mathrm{B-O}}$ the
distance between C$_\mathrm{B}$ and the overlayer atom Ir$_\mathrm{O}$, 
$d_{\mathrm{A-B}}$ the distance between two neighboring C atoms, and $\alpha$
the angle defined by the lines Ir$_\mathrm{S}$--C$_\mathrm{A}$
and C$_\mathrm{A}$--C$_\mathrm{B}$.
}
\label{tab:bonds}
\end{centering}
\end{table}

The situation is rather different when strong bonding between iridium and graphene 
in Ir-graphene-Ir structures occurs, at $z_\mathrm {Ir-g}$ around 2.3~{\AA} 
in Fig.~\ref{fig:vdW-DF}. First, we notice that the total energy depends strongly on
the position of the C atoms of graphene with respect to the Ir atoms below and 
above. Thus the two similar structures, (b) and (d) in Fig.~\ref{fig:sandwiches},
differ only in the position of the iridium atoms in the monoatomic overlayer, but the
total energies in Fig.~\ref{fig:LDA_PBE} differ by more than 1 eV! 
The absence of a stable strong bond in structure (b) shows that strong
bonding can occur only when both C atoms are saturated by Ir atoms, one directly
below and one above it. 
This immediately implies that the onset of strong bonding effectively anchors the 
iridium cluster and the underlying graphene to the particular spot of the 
moir\'e pattern of the graphene-substrate supercell.

The formation of the strong ``organometallic'' bond is accompanied by buckling of
graphene and C--C bond lengthening. Table~\ref{tab:bonds} shows the values
of bond lengths and angles corresponding to the strong bonding energy minima 
in the two panels of Fig.~\ref{fig:vdW-DF}.
These values are close to the ones of the tetrahedrally bonded C atoms in diamond,
and indicate that the rehybridization from $sp^2$ to $sp^3$ bonding has occurred, as
noted by Feibelman.~\cite{Feibelman2008}

\section{Discussion}
\label{sec:discussion}

These results show that the onset of the strong C--Ir binding in Ir-graphene-Ir 
sandwiches is accompanied by the 
disappearance of the aromatic character of the carbon
rings. The carbon atoms rehybridize to  $sp^3$ configuration, and 
the two C atoms in the graphene unit cell move
out of the plane in opposite directions. 
On the other hand, in one-sided
binding on Ir(111) (i.e. a clean graphene overlayer) the  C--Ir bond is
always weak, dominated by van der Waals interaction.

In order to get more insight into the character of the bonding of aromatic rings 
on Ir(111), we have made DFT calculations of benzene molecules C$_6$H$_6$ lying flat on 
the iridium surface. We found large differences of binding energies and distances 
for different positions of the benzene molecules with respect to the substrate atoms.
The most stable configuration is when the center of the ring is above a hollow
site, and the six C atoms are above three Ir atoms, two C on each Ir. (This 
configuration cannot be directly compared to any part of the moir\'e pattern
of graphene on Ir(111), Fig.~\ref{fig:moire}, since 
it corresponds to a different orientation of the aromatic rings, 
i.e. rotated by 30$^\circ$.) The C--Ir bond is around 2.4~{\AA}.
The GGA adsorption energy is somewhat less than 1~eV, and does not change substantially
when the vdW-DF nonlocal correlation is used.
The C atoms remain planar due to symmetry,
but the C--C bonds become longer, 1.43~{\AA} and 1.48~{\AA}, and the H atoms are 
slightly above the plane of the C atoms. This indicates a change of the nature of
the bonding of the carbon ring and a departure from the pure $sp^2$ hybridization.
The configurations where the center of the benzene ring is above an Ir atom,
and H atoms point either towards the neighboring Ir atoms or towards bridge sites,
are more weakly bound. 
The C--Ir bond length is around 3.4~{\AA}. 
In this case the nonlocal correlation is essential for the bonding. With
pure GGA functional there is virtually no bonding of the benzene molecule, 
while with the vdW-DF the bonding energy is around 0.6~eV.
The C--C bonds keep the value of 1.41~{\AA} as in the benzene molecule,
and the whole benzene structure is planar. These values are also very similar 
to those of graphene on Ir(111) obtained earlier. 
All these results indicate a weak, wan der Waals-dominated bonding.
Thus the bonding of benzene on Ir(111) shows even more variation of
bonding parameters than various regions of the moir\'e of graphene, 
indicating the richness of possible bonding of aromatic structures (molecules
and graphene) on metal surfaces.

Graphene strongly binds on some other surfaces, as mentioned in the 
Introduction,
apparently without strong rehybridization to  $sp^3$. 
In a recent combined experimental and theoretical
study of graphene bonding on Ru(0001) surface,~\cite{Ru0001Moritz}
it was found that graphene is strongly corrugated with a minimum C-Ru distance 
of 2.1~{\AA} 
and a corrugation of 1.53~{\AA}
in the regions of strong coupling.
The DFT calculations were performed using the standard PBE functional, which is
expected to work well in the regions of strong coupling, and the lack of the
van der Waals interaction which should be dominant in the weakly coupled 
`blisters' is not crucial.
The authors find that the height difference between neighboring C atoms 
in the graphene layer is below 0.03~{\AA} in the strong coupling region in DFT,
and conclude that the adsorbed graphene layer remains $sp^2$ hybridized.

Returning to the calculations of graphene on Ir(111), we note that
the use of the large supercell in Ref.~\onlinecite{Feibelman2008} 
has the advantage that
the lattice constants of both the iridium substrate and the graphene 
overlayer can be kept close to their natural values.
Thus the problems which we encounter
with our compressed Ir surface (expanded in the $z$ direction) are avoided.
In particular, it seems that 
we get somewhat
too large Ir-graphene distances compared to other calculations and the preliminary
experimental estimates.
Furthermore, when iridium clusters are added on top of graphene, in the large 
supercell approach the carbon 
atoms are free to relax both vertically and laterally, which is essential for
a good description of graphene buckling and the formation of the strong
C--Ir bond. In subsection~\ref{sub:vdW-DF} we had to use a 
calculational tour de force to detect the preference
of carbon atoms to lengthen somewhat the C--C bonds and thus approach more
closely the diamond structure. Furthermore, in the supercell approach the 
substrate iridium atoms may also relax laterally, optimizing the saturation
of the bonds to carbon atoms.

These advantages come with the downside that in 
Ref.~\onlinecite{Feibelman2008}
the  LDA functional was used. This was a necessary choice since GGA 
in the usual formulation, i.e. without the van
der Waals interaction being somehow accounted for, 
gives little or no binding of graphene (see Ref.~\onlinecite{NDiaye2006}, in
particular the Erratum).
However, LDA usually gives a too small equilibrium distance, and
overbinds in cases of strong chemical bonding. Thus in our 
calculations of graphite in
section~\ref{sec:graphite}, Fig.~\ref{fig:graphite},
LDA gave a too small interlayer distance. The LDA binding
energy of graphite was also too small since graphite is a system with very little
chemical component of the bond and the LDA overbinding could not compensate
fully the absence of the vdW component.

The compressed Ir(111) surface in our approach and the use of LDA in 
Ref.~\onlinecite{Feibelman2008} preclude a detailed quantitative comparison
between the results of the two calculations, and of each of them with experiment.
However, the semiquantitative agreement is good. 
Both approaches predict a rather weak bonding of a graphene monolayer with
the Ir(111) substrate, and the formation of a much stronger organometallic bond when 
iridium clusters are added on top, accompanied with the buckling of the graphene
structure and shortening of Ir--C distances.
For clean graphene, the Ir--C separation
at the 1-3 regions of the moir\'e pattern is around 3.48~{\AA} in 
Ref.~\onlinecite{Feibelman2008} and around 3.7~{\AA} 
in our work, while the 
experimental value has been estimated to around 3.38~{\AA}.
When the clusters trigger strong bonding and graphene buckling the Ir--C 
distance decreases to around 2.1~{\AA} in Ref.~\onlinecite{Feibelman2008} 
and to around 2.2~{\AA} in our work, while the Ir--C--C angles are around 105$^\circ$.

The bonding of graphene on some other (111) surfaces of fcc metals
assuming commensurate configurations has also been investigated. 
In the papers by Giovannetti et al.~\cite{Giovannetti2008} and
Khomyakov et al.,~\cite{Khomyakov2009} the LDA functional was used.
The unit cells were either 2 graphene C atoms and one metal atom in each 
layer (e.g. Ni, Co, Cu), as in our calculation, or 8 C atoms and 3 metal atoms with
the graphene unit cell rotated by ${30^\circ}$ when the difference of the 
lattice constants was larger (e.g. Pd, Au, Pt). It was found that 
graphene interacts strongly with Ni, Co, and Pd, with the equilibrium 
metal-graphene distance between 2.05 and 2.30~{\AA}, and weakly with 
Cu, Au and Pt, with equilibrium distance between 3.26 and 3.31~{\AA}. 
These findings are in agreement with experiment, where available.
The mismatch of the lattice constant of graphene is
4\% for Cu, 1.2\% for Ni and 2\% for Co (metal unit cell
being larger in all three cases). This is clearly smaller than 
in our calculation, where the difference of Ir(111) and graphene 
lattice constants is around 10\%.

Vanin et al.~\cite{Vanin} consider the same surfaces, but in a quite different 
approach. They use the vdW-DF correlation functional~\cite{Dion2005}
evaluated using the method proposed in Ref.~\onlinecite{Soler} and
self-consistently implemented into the real-space projector augmented wave
gpaw code.~\cite{gpaw} They do not adjust the metal substrate to match the
lattice constant of graphene, but instead keep it at their experimental lattice
parameters and adjust the graphene sheet. They claim that the vdW-DF results
do not change significantly if they fix the graphene lattice
parameter to its optimized value and adjust the metals correspondingly.
Surprisingly, they obtain weak binding for all metals considered, with 
metal-graphene distances in the range 3.40--3.72~{\AA}. This is in clear
disagreement for Co and Ni, where strong binding has been experimentally
confirmed.

In contrast to this, our calculations of Ir(111)-graphene structures
give
an overall agreement with
other calculations and with experiment.
The weak vdW
minimum around 3.7~{\AA} which exist both for clean graphene
overlayer (Fig.~\ref{fig:vdW-DF-1}) and for graphene with iridium adclusters
(Fig.~\ref{fig:vdW-DF}) is obtained correctly only with vdW-DF.
The overall shape of the potential minimum of the strong bond around 
2.3~{\AA} for iridium adclusters is roughly similar for calculations using
LDA and PBE with vdW-DF, although LDA clearly overbinds. Even plain
PBE calculations show a comparable local minimum, just weaker.
The disagreement found in Ref.~\onlinecite{Vanin} is therefore even more 
surprising. We have not tried our method on metals other than iridium,
where the strong chemisorption minimum exists only if adclusters are 
present. There is a possibility that vdW-DF 
does not work so well for other metals.
However, in our opinion the source of disagreement may also be the other 
approximations used in Ref.~\onlinecite{Vanin}.
 
The graphene is particularly stable due to the aromatic character
of the carbon rings. Perturbing the structure (for example by forcibly 
changing the natural bond length) may significantly change the reactivity. 
Thus simply adapting the
graphene lattice constant to the substrate may have unwanted consequences,
weakening the stability of the aromatic bonds, as well as changing the doping 
of the graphene layer in contact with the metal surface.
The opposite procedure, which we used in this paper, i.e. adapting the 
substrate lattice constant, seems preferable to us but 
may also have some weaknesses. First, the 
change of the electronic structure of the substrate may be large enough to 
alter the reactivity compared to the natural metal. Also, the lattice
constant of the free graphene may not be optimal for rehybridized graphene 
forming strong $sp^3$ bonds.
We had to expand the graphene lattice slightly in order to
obtain a sufficiently stable strong bonding.
In the process we had to carefully 
account for the change in energy of the iridium substrate, which was,
of course, also expanded (i.e. less compressed compared to the
natural structure).
All this indicates that the graphene lattice constant should be left at 
its natural value in the weak bonding cases, but should be allowed to
relax and lengthen when the strong bonding regime accompanied with 
graphene buckling and rehybridization to diamondlike bond occurs.
This cannot be achieved in the simplified commensurable 
geometries, and a full large supercell calculation with state-of-the-art
nonlocal correlation functional seems to be the only approach which
can give the answer about the structure of graphene adsorbed on
various metals 
in the general case.

\section{Conclusions}
\label{sec:conclusions}

We find that a graphene monolayer on Ir(111) is weakly bound, and keeps
the aromatic character of the carbon rings.
In Ir-graphene-Ir structures C atoms show a tendency towards rehybridization
and formation of $sp^3$ 
bonds, which in favorable cases (an Ir atom directly below or above each C atom) are
more stable that the physisorbed structure.
In all cases, the use of the vdW-DF which includes a full description of the 
nonlocal correlation is essential.
However, our approach in which the substrate lattice constant is adjusted to
match graphene does not give full quantitative accuracy. In order to
obtain that kind of agreement, large calculations on
realistic supercells using DFT functionals with nonlocal correlation are 
necessary. This conclusion is also true for other graphene-on metal systems, 
in which the nature of the graphene-metal bond may be quite different than
on Ir(111).

\begin{acknowledgments}
This work was supported by the Ministry of Science, Education and Sports 
of the Republic of Croatia, under Contract No.~098-0352828-2863.
P. Lazi\'c acknowledges the financial support from
Alexander von Humboldt foundation.
\end{acknowledgments}


\begin{thebibliography}{11}
\bibitem{123}
   The regions which we denote by 1-3, 1-2, and 2-3, are called `hcp'',
      ``fcc'', and ``atop'' regions in Refs.~\onlinecite{NDiaye2006}
      and \onlinecite{Feibelman2008}, according to the Ir(111) 
      high-symmetry site which
      is surrounded by C-atom rings, i.e. which is {\em not} covered by a
      graphene atom.
\bibitem{ABnotation}
   The labels A and B for the two inequivalent sublattices of graphene is not to
      be confused with the same letters used to denote the graphene stacking in 
      Section~\ref{sec:graphite}. 
      Both notations are widely used.
\end{thebibliography}
\end{document}